\documentclass[aps,12pt,showkeys,showpacs]{revtex4}
\usepackage{psfig,epsfig}

\newcommand{\bm}{\bibitem}
\newcommand{\cp}{\chi^{(+)}}
\newcommand{\cm}{\chi^{(-)*}}

\newcommand{\vv}{V_{bc}({\bf r}_1)}
\newcommand{\ri}{{\bf r}_i}
\newcommand{\ro}{{\bf r}_1}
\newcommand{\ak}{{\bf k}_a}
\newcommand{\bq}{{\bf k}_b}

\newcommand{\rc}{{\bf r}_c}
\newcommand{\cq}{{\bf k}_c}

\begin{document}


\title{A full quantal theory of one-neutron halo breakup reactions}
\date{\today}
\author{R. Chatterjee}
\email{raja@theory.saha.ernet.in}
\affiliation{
Theory Group, Saha Institute of Nuclear Physics,
1/AF, Bidhannagar, Kolkata 700064, India.}

\begin{abstract}
We present a theory of one-neutron halo breakup reactions within the framework
of post-form distorted wave Born approximation 
wherein pure Coulomb, pure nuclear and
their interference terms are treated consistently in a single setup. 
This formalism is used to study the breakup of one-neutron halo
nucleus $^{11}$Be on several
targets of different masses. We investigate the role played by the
pure Coulomb, pure nuclear and
the Coulomb-nuclear interference terms by calculating several reaction
observables. The Coulomb-nuclear interference terms are found to be important
for more exclusive observables.
\end{abstract}
\maketitle

\section{Introduction}
Our current understanding of most of the processes governing the
nuclear systems is based on studies made with stable nuclei, 
which constitute less than ten percent of all the nuclei known to exist 
in nature. Away from the valley of stability there are a large number
of nuclei having very short half lives and very small one- and two-nucleon
separation energies. Many of them exhibit a halo structure
in their ground states in which loosely bound valence nucleon(s)
has (have) a large spatial extension with respect to the respective
core \cite{han01,tani,rii94,bert93}. 
We still lack a fully microscopic understanding
of the stability of these unique many body systems.

These nuclei are also important from nuclear astrophysics point of view.
The r-process and the s-process paths which together are dominant mechanisms
for nucleosynthesis of heavy elements above iron,  pass through the 
region of neutron rich exotic nuclei, in the Segr\'{e} chart. 
Properties of these nuclei are, therefore, important inputs to 
theoretical calculations of stellar burning 
which otherwise are often forced to rely on global assumptions about
nuclear masses, decays and level structures extracted from stable nuclei.

Projectile breakup reactions have played a major role in unraveling the
structure and properties of halo nuclei. 
However it is clear
that pure Coulomb \cite{nak94,nak99,pb1,pb2,cha00,cha01,pbc} 
or pure nuclear \cite{bhe,bon00,yab92}
breakup calculations may not be fully sufficient to describe all the 
details of the
halo breakup data which have been increasing rapidly both in quality and 
quantity \cite{han95,dav98,gil01,ritu02,udp}. In majority of them both 
Coulomb and nuclear breakup effects as well as their interference terms are
likely to be significant and the neglect of the latter terms may not
be justified \cite{ann90,ann94,pb93,mad01}. A theory which can
take care of the Coulomb and nuclear breakup effects as well as 
their interference terms on an equal footing is an important
requirement in interpreting the data obtained from the experiments 
done already or are planned to be done in future.

For breakup reactions of light stable nuclei, such a theory has been
developed \cite{bau84} within the framework of post-form distorted
wave Born approximation (DWBA), which successfully describes the 
corresponding data at low beam energies. However, since it uses the 
simplifying approximation of a zero-range interaction \cite{sat64}
between constituents of the projectile, it is inapplicable to cases
where the internal orbital angular momentum of the projectile is 
different from zero.  

Recently, we have presented a theory \cite{cha02,thesis} to 
describe the breakup 
reactions of one-nucleon halo nuclei within the post-form DWBA (PFDWBA)
framework, that  
consistently includes both Coulomb and nuclear interactions between the
projectile fragments and the targets to all orders, but treats the 
fragment-fragment interaction in first order. The Coulomb and nuclear
breakups along with their interference term are treated within a
single setup in this theory. The breakup contributions from the
entire continuum corresponding to all the multipoles 
and the relative orbital angular momenta between the valence
nucleon and the core fragment are included in this theory where
finite range effects are treated by a local momentum approximation
(LMA) \cite{shy85,bra74}. Full ground state wavefunction of the
projectile, of any angular momentum structure, enters as an
input to this theory.

The Coulomb-nuclear interference (CNI) terms have also
been calculated using the prior-form DWBA \cite{shy99} and within 
models \cite{typ01,bon02} where the time evolution of the 
projectile in coordinate space is described by solving the 
time dependent Schr\"{o}dinger equation, treating
the projectile-target (both Coulomb and nuclear) interaction as 
a time dependent external perturbation. 
Recently, within an eikonal-like framework, Coulomb and nuclear 
processes have also been treated within the same framework
in Ref. \cite{bon03}.

In this paper, we present more details of the formalism of 
PFDWBA breakup theory and of its application to  
breakup reactions of the one-neutron halo nucleus $^{11}$Be on targets of
masses spanning a wide range in the periodic table. We investigate
the role played by the pure Coulomb, pure nuclear and the CNI terms
by calculating different breakup observables. Our formalism is presented 
in section II. In section III, we present and discuss the results of our 
calculations for various observables for the breakup of $^{11}$Be on
various targets. Summary and conclusions of our work  are presented
in section IV. Additional discussions on the validity of the LMA  relevant
for our case are presented in Appendix A. 

\section{Formalism}

We consider the elastic breakup reaction, $a + t \to b + c + t$, in which the
projectile $a$ ($a = b +c$) breaks up into fragments $b$ and $c$ (both of
which can be charged) in the Coulomb and nuclear fields of a target $t$.
The triple differential cross section for this reaction is given by
\begin{eqnarray}
{{d^3\sigma}\over{dE_bd\Omega_bd\Omega_c}} & = &
{2\pi\over{\hbar v_a}}\rho(E_b,\Omega_b,\Omega_c)
\sum_{\ell m}|\beta_{\ell m}|^2, \label{eq1_5}
\end{eqnarray}
where $v_a$ is the relative velocity of the projectile with 
respect to the target, $\ell$ is the orbital angular momentum for the 
relative motion of $b$ and $c$ in the ground state of $a$, and 
$\rho(E_b,\Omega_b,\Omega_c)$ is the appropriate phase space factor (see,
e.g., Ref. \cite{cha00}).
The reduced transition amplitude, in Eq. (\ref{eq1_5}), $\beta_{\ell m}$ is
defined as 
\begin{eqnarray}
\hat{\ell}\beta_{\ell m}(\bq,\cq;\ak) & = & 
\int d\ro d\ri\cm_b(\bq,{\bf r})\cm_c(\cq,\rc) \vv \nonumber \\
& & \times u_\ell (r_1) Y^\ell_{m} ({\hat r}_1)\cp_a(\ak,\ri), \label{eq2_5}
\end{eqnarray}
with ${\hat \ell} \equiv \sqrt{2\ell + 1}$.
In Eq. (\ref{eq2_5}), functions $\chi_i$ represent 
the distorted waves for the relative motions of various particles 
in their respective channels with appropriate 
boundary conditions. 
The superscripts $(+)$  and $(-)$ represents 
outgoing and ingoing wave boundary conditions, respectively.
Arguments of these functions contain the
corresponding Jacobi momenta and coordinates.  $\vv$ represents the
interaction between $b$ and $c$, and $u_\ell (r_1)$ is the radial part of
the corresponding wavefunction in the ground state of $a$. 
The position vectors satisfy the relations 
(see also Fig. 1 of Ref. \cite{cha00}): 
\begin{eqnarray}
{\bf r} &=& \ri - \alpha\ro,~~ \alpha = {m_c\over {m_c+m_b}},   \\
\rc &=& \gamma\ro +\delta\ri, ~~ \delta = {m_t\over {m_b+m_t}}, 
~~  \gamma = (1 - \alpha\delta),    
\end{eqnarray}
where $m_i$ ($i=a,b,c,t$) are the masses of various particles.   

The reduced amplitude $\beta_{\ell m}$ [Eq. (\ref{eq2_5})] involves 
a six-dimensional integral which makes its evaluation quite complicated.
The problem gets further aggravated due to the fact that the integrand
involves the product of three scattering waves that exhibit an oscillatory
behavior asymptotically. In order  
to facilitate an easier computation of Eq.~(\ref{eq2_5}), 
we perform a Taylor series expansion of the distorted waves
of particles $b$ and $c$ about ${\bf r}_i$ and write 
\begin{eqnarray}
\chi^{(-)}_b(\bq,{\bf r}) & = & e^{-i\alpha{\bf K}_b.\ro}
                           \chi^{(-)}_b(\bq,\ri), \label{eq3_5} \\
\chi^{(-)}_c(\cq,{\bf r}_c) & = & e^{i\gamma{\bf K}_c.\ro}
                           \chi^{(-)}_c(\cq,\delta\ri).  \label{eq4_5}
\end{eqnarray}
Employing the LMA \cite{shy85,bra74}, the magnitudes
of momenta ${\bf K}_j$ are taken as 
\begin{eqnarray}
K_j(R)  = \sqrt {(2m_j/ \hbar^2)[E_j- V_j(R)]}, \label{ea6}
\end{eqnarray}
 where $m_j$ ($j=b,c$) is the reduced mass of the $j-t$ system,
$E_j$ is the energy of particle $j$ relative to the target in the
center of mass (c.m.) system, and $V_j(R)$ is the potential between $j$ and 
$t$ at a distance $R$. Substituting
Eqs. (\ref{eq3_5}) and (\ref{eq4_5}) in Eq. (\ref{eq2_5}), 
 the amplitude $\beta_{\ell m_\ell}$ factorizes into
two terms, each involving a three-dimensional integral,
\begin{eqnarray}
{\hat \ell} \beta_{\ell m_\ell} = I_f \times I,  \label{eq5_5}
\end{eqnarray}
where 
\begin{eqnarray}
I_f = \int d{\bf r}_1 e^{-i{\bf Q}.{\bf r}_1} 
 V_{bc}({\bf r}_1)u_\ell(r_1) Y^l_{m_\ell}({\bf r}_1),  \label{eq6_5}
\end{eqnarray}
with 
\begin{eqnarray}
{\bf Q} = \gamma {\bf K}_c - \alpha {\bf K}_b, \label{eqQ_5}
\end{eqnarray}
 and
\begin{eqnarray}
I = \int d{\bf r}_i \chi_b^{(+)}(-{\bf k}_b,{\bf r}_i) 
\chi_c^{(+)}(-{\bf k}_c,\delta {\bf r}_i)
\chi_a^{(+)}({\bf k}_a,{\bf r}_i), \label{eq7_5}
\end{eqnarray}
where we have used the relation 
$\chi^{(-)*}({\bf k},{\bf r}) = \chi^{(+)}(-{\bf k},{\bf r})$.

Let us consider first the integral $I$. We expand
the distorted wave for projectile-target relative motion in partial waves as
\begin{eqnarray}
\chi_a^{(+)}({\bf k}_a,{\bf r}_i) = {{4\pi}\over {k_a r_i}} 
\sum_{L_aM_a} i^{L_a}
f_{L_a}(k_a,r_i) Y^{L_a*}_{M_a}({\hat {\bf k}}_a)
Y^{L_a}_{M_a}({\hat {\bf r}}_i), \label{eq7a_5}
\end{eqnarray}
where $f_{L_a}(k_a,r_i)$ is the radial part, calculated by solving the 
Schr\"{o}dinger equation with proper optical potentials, which includes
both Coulomb and nuclear terms. Beyond the range of the nuclear potential
$f_{L_a}(k_a,r_i)$ has the form
\begin{eqnarray}
f_{L_a}(k_a,r_i) \stackrel{r_i \to \infty} = {i\over 2} e^{i\sigma_{L_a}}
[H^{(-)}_{L_a}(k_ar_i) - \epsilon_{L_a} H^{(+)}_{L_a}(k_ar_i)], \label{eq8_5}
\end{eqnarray}
where $H^{(\pm)}_L = G_L \pm iF_L$, with $F_L$ and $G_L$ being the regular
and irregular Coulomb wavefunctions, respectively, and $\epsilon_L$ is 
the scattering phase shift of the $L^{\rm th}$ partial wave. In
Eq. (\ref{eq8_5}), $\sigma_L = {\rm arg}\Gamma(L+1+i\eta)$
is the Coulomb phase shift with Coulomb parameter $\eta$. 

Similar expansions can be written for the distorted waves for the 
core fragment and the valence particle relative motions:
\begin{eqnarray}
\chi_b^{(+)}(-{\bf k}_b,{\bf r}_i) &=&  
{{4\pi}\over {k_b r_i}} 
\sum_{L_bM_b} i^{-L_b} f_{L_b}(k_b,r_i) 
Y^{L_b}_{M_b}({\hat {\bf k}}_b)Y^{L_b*}_{M_b}({\hat {\bf r}}_i)
\label{eq9_5}
\end{eqnarray}
\begin{eqnarray}
\chi_c^{(+)}(-{\bf k}_c,\delta {\bf r}_i) = {{4\pi}\over {k_c\delta r_i}} 
\sum_{L_cM_c} i^{-L_c} f_{L_c}(k_c,\delta r_i) 
Y^{L_c}_{M_c}({\hat {\bf k}}_c)Y^{L_c*}_{M_c}({\hat {\bf r}}_i) \label{eq10_5}
\end{eqnarray}
If ${\hat {\bf k}}_a$(the incident beam direction) is chosen along the 
$\hat z$-direction, then the spherical harmonic, 
$Y^{L_a*}_{M_a}({\hat {\bf k}}_a)$, in 
Eq. (\ref{eq7a_5}) simplifies to
\begin{eqnarray}
 Y^{L_a*}_{M_a}(\theta=0,\phi=0) = 
        {{{\hat L}_a}\over {\sqrt {4\pi}}} \delta_{M_a,0} \nonumber\\
 \Rightarrow M_a = 0 \Rightarrow M_b=-M_c = M ~ ({\rm say}).
\end{eqnarray}
Thereafter, 
substituting Eqs. (\ref{eq7a_5}), (\ref{eq9_5}) and (\ref{eq10_5}) 
in Eq. (\ref{eq7_5}), we obtain
\begin{eqnarray}
I &=& {(4\pi)^2 \over {k_a k_b k_c\delta }} \sum_{L_aL_bL_cM} 
 (-)^M (i)^{L_a-L_b-L_c} {\hat L}_b{\hat L}_c
   Y^{L_b}_{M}({\hat {\bf k}}_b)Y^{L_c*}_{M}({\hat {\bf k}}_c)  \nonumber\\
&\times& \langle L_b M L_c -M|L_a 0 \rangle \langle L_b 0 L_c 0| L_a 0 \rangle 
     \nonumber\\
&\times & \int_0^{\infty} {{dr_i}\over {r_i}} 
f_{L_a}(k_a,r_i) f_{L_b}(k_b,r_i) f_{L_c}(k_c,\delta r_i). \label{eq14_5}
\end{eqnarray}
 
Let us now turn our attention to integral $I_f$ 
[Eq. (\ref{eq6_5})], which contains the structure information.
Expanding the exponential, in Eq. (\ref{eq6_5}), in partial waves,
$I_f$ simplifies to
\begin{eqnarray}
I_f = 4\pi i^{-\ell} Y^\ell_{m_\ell}({\hat {\bf Q}}) 
\int_0^{\infty} r_1^2 dr_1 j_{\ell}(Qr_1) u_\ell(r_1)V_{bc}(r_1). \label{eq16_5} 
\end{eqnarray}
Substituting Eqs. (\ref{eq14_5}) and (\ref{eq16_5}) in Eq. (\ref{eq5_5}),
we get
\begin{eqnarray}
{\hat \ell} \beta_{\ell m} &=&
{(4\pi)^{3} \over {k_a k_b k_c\delta }} i^{-\ell} Y^\ell_{m_\ell}({\hat {\bf Q}})
Z_\ell (Q) \sum_{L_aL_bL_c} (i)^{L_a-L_b-L_c} {\hat L}_b{\hat L}_c
\nonumber \\
& \times & {\cal Y}^{L_b}_{L_c}({\hat k}_b,{\hat k}_c)
 \langle L_b 0 L_c 0| L_a 0 \rangle 
 {\cal R}_{L_b,L_c,L_a}(k_a,k_b,k_a), \label{eq17_5}
\end{eqnarray}
where
\begin{eqnarray}
{\cal Y}^{L_b}_{L_c}({\hat k}_b,{\hat k}_c) & = &
\sum_M (-)^M\langle L_b M L_c -M|L_a 0 \rangle
Y^{L_b}_{M}({\hat {k}}_b)Y^{L_c*}_{M}({\hat {k}}_c), \\  
Z_\ell (Q) & = & \int_0^{\infty} r_1^2 dr_1 j_{\ell}(Qr_1) u_\ell(r_1)V_{bc}(r_1),
 \\ 
{\cal R}_{L_b,L_c,L_a} & = & \int_0^{\infty}
{{dr_i}\over {r_i}} f_{L_a}(k_a,r_i)
f_{L_b}(k_b,r_i) f_{L_c}(k_c,\delta r_i). \label{eq18_5} 
\end{eqnarray}
The slowly converging radial integral ${\cal R}_{L_b,L_c,L_a}$ 
[Eq. (\ref{eq18_5})] can be
effectively handled by using the complex plane method~\cite{vin70,thesis}. 
$Y^\ell_{m_\ell}({\hat {\bf Q}})$, in Eq. (\ref{eq17_5}), 
whose argument contains the direction of a vector which is the 
sum of two other vectors [Eq. (\ref{eqQ_5})], can be expressed in 
terms of spherical harmonics corresponding to directions of
 those vectors as \cite{moshin}
\begin{eqnarray}
(|{\bf Q}|)^\ell Y^\ell_{m_\ell}({\hat {\bf Q}}) &=&
  \sum_{L M_L} \frac{\sqrt{4\pi}}{\hat{\ell}}
  {\left(\begin{array}{c}2\ell +1 \\
   2L\end{array}\right)}^{1/2} 
   (|\alpha K_b|)^{\ell - L} (\gamma K_c)^{L} \nonumber \\
& \times & \langle \ell- L m_\ell - M_L L M_L | \ell m_\ell \rangle \nonumber \\ 
& \times & Y^{\ell - L}_{m_\ell - M_L}({\hat {\bf K}}_b) 
            Y^L_{M_L}({\hat {\bf K}}_c),
\end{eqnarray}
where $L$ runs from $0$ to $\ell$ and
\begin{eqnarray}
{\left(\begin{array}{c}x \\
   y\end{array}\right)} = \frac{x!}{y! (x-y)!}
\end{eqnarray}
is the binomial coefficient.

This theory can be used to calculate breakup of both neutron
and proton halo nuclei. Generally, the maximum value of the partial
waves $L_a,L_b,L_c$ must be very large in order to ensure the convergence
of the partial wave summations in Eq.~(\ref{eq17_5}). However, for 
the case of
one-neutron halo nuclei, one can make use of the following method 
to include summations over infinite number of partial
waves. We write $\beta_{\ell m}$ as
\begin{eqnarray}
\beta_{\ell m} & = & \sum_{L_i = 0}^{L_{i}^{max}} {\hat \beta}_{\ell m} (L_i) 
             + \sum_{L_i = L_{i}^{max}}^{\infty} 
                    {\hat \beta}_{\ell m}(L_i), \label{eq19_5}   
\end{eqnarray}
where ${\hat \beta}$ is defined in the same way as Eq.~(\ref{eq17_5}) 
except for
the summation sign and $L_i$ corresponds to $L_a$, $L_b$, and $L_c$. If
the value of $L_i^{max}$ is chosen to be appropriately large, the
contribution of the nuclear field to the second term of Eq.~(\ref{eq19_5}) 
can be
neglected and we can write
\begin{eqnarray}
\sum_{L_i = L_{i}^{max}}^{\infty}{\hat \beta}_{\ell m}(L_i)  \approx    
            \sum_{L_i = 0}^{\infty}{\hat \beta}_{\ell m}^{Coul}(L_i) -   
             \sum_{L_i = 0}^{L_{i}^{max}}{\hat \beta}_{\ell m}^{Coul} (L_i),
\end{eqnarray}
where the first term on the right hand side, is the pure Coulomb 
breakup amplitude which for
the case where one of the outgoing fragments
is uncharged, can be expressed analytically in terms of the 
bremsstrahlung integral (see Ref. \cite{cha00}). Therefore, only
two terms, with reasonable upper limits, are required to be evaluated
by the partial wave expansion in Eq.~(\ref{eq19_5}).

\section{Calculations on $^{11}$Be}

\subsection{Structure model and optical potentials}
The wavefunction, $u_\ell(r)$, appearing in the structure term, $Z_\ell$,
has been calculated by adopting a single particle potential
model in the same way as in Ref. \cite{cha00}. 
The ground state of $^{11}$Be was considered to be a predominantly $s$-state
with a $2s_{1/2}$ valence neutron coupled to the $0^+$ $^{10}$Be core
[$^{10}$Be $\otimes$ $2s_{1/2}\nu$] with a one-neutron separation energy 
of 504 keV and a spectroscopic factor of 0.74 \cite{aum}.
The single particle wavefunction was constructed by assuming the 
valence neutron-$^{10}$Be interaction to be of Woods-Saxon type
whose depth was adjusted to reproduce the corresponding value
of the binding energy with fixed values of the radius and diffuseness
parameters (taken to be 1.15 fm and 0.5 fm, respectively). 
This gave a potential depth of 71.03 MeV, a root mean square (rms) radius
for the valence neutron of 6.7 fm, and a rms radius for $^{11}$Be
of 2.91 fm when the size of the $^{10}$Be core was taken to be 2.28 fm.
 The neutron-target optical potentials used by us 
were extracted from the global set of Bechhetti-Greenlees
(see, e.g,~\cite{per76}), while those used for the $^{10}$Be-target
(\cite{per76,beu98}) system are shown in Table I. Following~\cite{typ01},
we have used the sum of these two potentials for the
$^{11}$Be-target channel. We found that values of $L_i^{max}$ of 500
for Au, Ta, U, Pb and Ti targets and 150 for 
Be and C targets provided very good convergence of the
corresponding partial wave expansion series [Eq.~(\ref{eq17_5})]. The 
local momentum wave vectors are evaluated at a distance, $R$ = 10 fm
in all the cases,
and their directions are taken to be same as that of asymptotic momenta
(see appendix A).
\begin{table}[h]
\caption{Optical potential parameters for the $^{10}$Be-target
interaction. Radii are calculated with the $r_jt^{1/3}$ convention.}
\begin{center}
\begin{tabular}{|c|c|c|c|c|c|c|}
\hline
system &  $V_r$ & $r_r$& $a_r$&  $W_i$& $r_i$& $a_i$\\
       & (MeV) & (fm) & (fm) & (MeV) & (fm) & (fm)      \\       
\hline
$^{10}$Be--$^{197}$Au & 400 & 2.08 & 0.9 & 76.2 & 1.52 & 0.38  \\
$^{10}$Be--$^{208}$Pb & 400 & 2.08 & 0.9 & 76.2 & 1.52 & 0.38  \\
$^{10}$Be--$^{44}$Ti &70 & 2.5 & 0.5 & 10.0 & 1.5 & 0.50 \\
$^{10}$Be--$^{9}$Be & 100 & 2.6 & 0.5& 18.0 & 2.6 & 0.50  \\
\hline
\end{tabular}
\end{center}
\end{table}

\subsection{Neutron energy distribution}
\begin{figure}[ht]
\begin{center}
\mbox{\epsfig{file=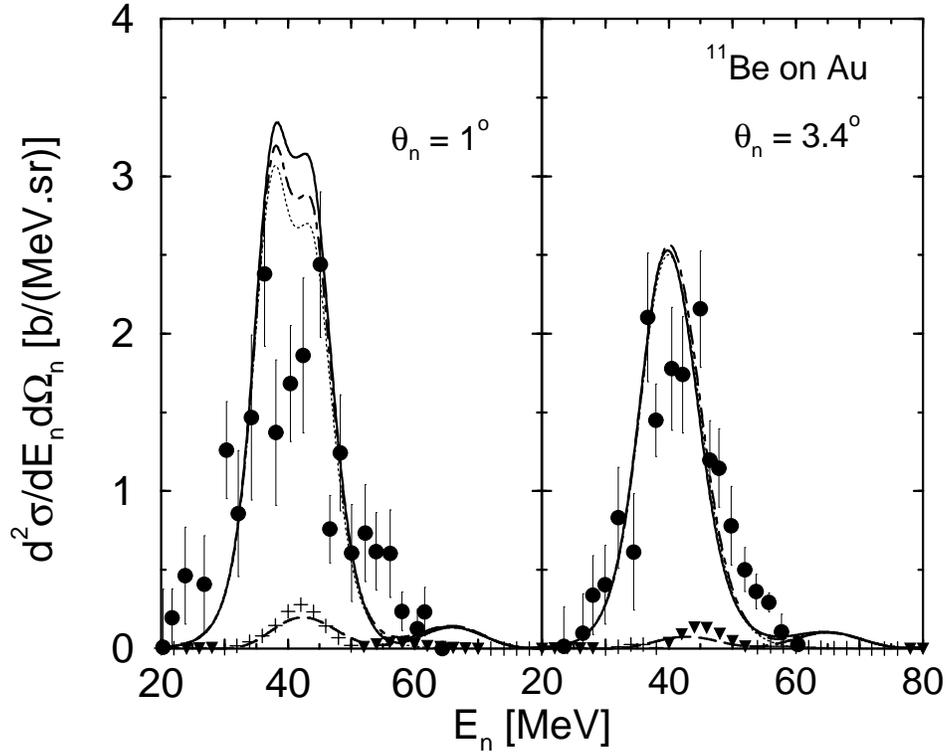,width=0.75\textwidth}}
\end{center}
\vskip .1in
\caption {
Neutron energy distribution for the breakup reaction $^{11}$Be on Au 
at the beam energy of 41 MeV/nucleon, at the neutron angles of
$1^{\circ}$ and $3.4^{\circ}$. 
The dotted and dashed
lines represent the pure Coulomb and nuclear contributions, respectively
while their coherent and incoherent sums are shown by the solid 
and dot-dashed lines, respectively. The plus signs
and the inverted solid triangles represent the magnitudes of the 
positive and negative interference terms, respectively. The data are
taken from \cite{ann94}. 
}
\label{fig:figb_5}
\end{figure}

In Fig.\ 1, we present the results of our calculations for 
the double differential cross section as a function of 
neutron energy for two neutron angles ($1^{\circ}$ and $3.4^{\circ}$),
in the breakup of $^{11}$Be on Au at the beam energy of 41 MeV/nucleon.
The core scattering angle in the laboratory system has been integrated from
$0^{\circ}$ to $30^{\circ}$.
The dotted and dashed
lines represent the pure Coulomb and pure nuclear contributions, respectively,
while their coherent and incoherent sums are shown by the solid 
and dot-dashed lines, respectively. The plus signs
and the inverted solid triangles represent the magnitudes of the 
positive and negative interference terms, respectively.

The CNI terms are seen to be dependent on angles and energies of 
the outgoing neutron. Their
magnitudes are nearly equal to those of the nuclear breakup contributions
which leads to a difference in the incoherent and coherent sums of the
Coulomb and nuclear contributions underlying thus the importance of
these terms.

\subsection{Relative energy spectra}

\begin{figure}[ht]
\begin{center}
\mbox{\epsfig{file=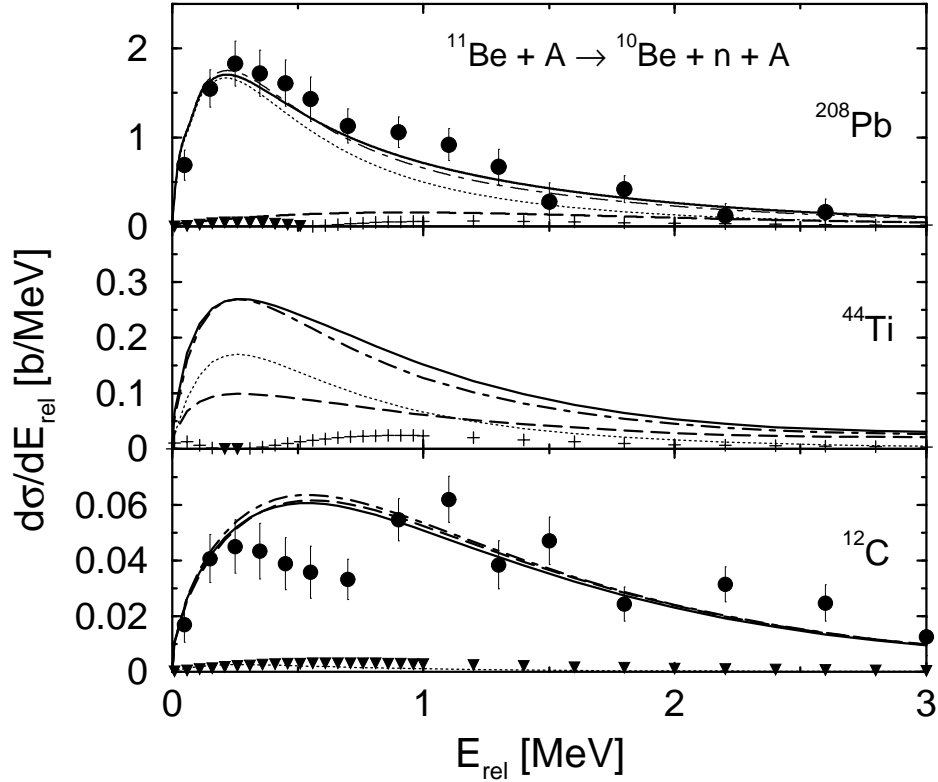,width=0.75\textwidth}}
\end{center}
\caption {
The differential cross section as a function of the relative energy of the
fragments (neutron and $^{10}$Be) in the breakup reaction of $^{11}$Be
on $^{208}$Pb, $^{44}$Ti and $^{12}$C targets at 72 MeV/nucleon. 
The dotted and dashed
lines represent the pure Coulomb and nuclear breakup 
contributions, respectively while their coherent and incoherent sums are 
shown by the solid and dot-dashed lines, respectively. 
The plus signs and the inverted triangles represent the magnitudes of the 
positive and negative interference terms, respectively. 
The data are taken from~\protect\cite{nak94}. 
}
\label{fig:figc_5}
\end{figure}
The relative energy spectrum of the fragments (neutron and $^{10}$Be) 
emitted in the breakup of $^{11}$Be on   
$^{208}$Pb (top panel), $^{44}$Ti (middle panel) and 
$^{12}$C (bottom panel) targets at the beam energy of 72 MeV/nucleon
is shown in Fig. 2. 
In these calculations the integration over the projectile c.m. angle
($\theta_{n{^{10}{\rm Be}}-{\rm Pb}}$) has been done in the range of 
$0^{\circ}$--$40^{\circ}$, mainly to include the effects of 
nuclear breakup coming from small impact parameters. The relative
angle between the fragments ($\theta_{n-{^{10}{\rm Be}}}$) 
has been integrated from $0^{\circ}$ to $180^{\circ}$.
The dotted and dashed
lines represent the pure Coulomb and nuclear breakup 
contributions, respectively while their coherent and incoherent sums are 
shown by the solid and dot-dashed lines, respectively. 
The plus signs and the inverted triangles represent the magnitudes of the 
positive and negative interference terms, respectively. 

In case of breakup on a heavy target ($^{208}$Pb) [Fig. 2 (top panel)] the pure
Coulomb contributions dominate the cross sections around the peak value,
while at larger relative energies
the nuclear breakup is important.
This is attributed to the different energy dependence of the two
contributions~\cite{typ01}.
The nuclear breakup occurs when the projectile and the 
target nuclei are close to each other. Its magnitude, which is determined
mostly by the geometrical conditions, has a weak dependence
on the relative energy of the outgoing fragments beyond a certain
minimum value. In contrast, the Coulomb breakup contribution has a
long range and it shows a strong energy dependence. 
The number of virtual photons increases for small excitation energies
and hence the cross sections
rise sharply at low excitation energies. After a certain value of this
energy the cross sections decrease due to setting in of the
adiabatic cut-off.
The coherent sum of the Coulomb and nuclear contributions provides a
good overall description of the experimental data. The nuclear and the
CNI terms are necessary to explain the data at larger relative energies.

In the middle panel of Fig. 2, we show the relative energy of the fragments
in the breakup of $^{11}$Be on a medium mass target ($^{44}$Ti). At low
relative energies the pure Coulomb contributions are slightly higher than
the pure nuclear ones, while at higher relative energies it is the nuclear
part which dominates. Apart from the very low relative energy region the 
CNI terms play an important role, which is clearly borne out by the
difference in the coherent (solid) and incoherent (dot-dashed) sums of 
the pure Coulomb and pure nuclear contributions.

The relative energy spectra for the breakup on a light target ($^{12}$C)
is shown in the bottom panel of Fig. 2. In this case we have used the
same optical potential for the $^{10}$Be-$^{12}$C system as in the
$^{10}$Be-$^{9}$Be case, which we had used earlier in calculating the
neutron angular distribution in Ref. \cite{cha02}. The total cross section
in this case is normalized to the experimental cross section (found by 
integrating the area under the data points) and the same normalization 
constant is used for all the cross sections in this case.
The breakup is clearly seen to be
nuclear dominated at all relative energies, and the pure Coulomb and CNI terms
have very little contributions.

The importance of the
peripheral region even in nuclear dominated reactions
is underlined in Fig. 3. In this figure, we show a comparison
of the angular distribution of the single neutron observed in the
elastic breakup of $^{11}$Be on a $^{9}$Be target at the beam energy of 41
MeV/nucleon, calculated with (dashed line) and without (solid line)
a lower cut-off of 10 fm in the $r_i$ integral [Eq. (\ref{eq18_5})]. The close
similarity of the two results shows the importance of the peripheral
region in this reaction. This fact is further strengthened by noting that
the total one-neutron removal cross sections calculated with and
without cut-off are found to be 0.128 b and 0.189 b, respectively.
\begin{figure}[ht]
\begin{center}
\mbox{\epsfig{file=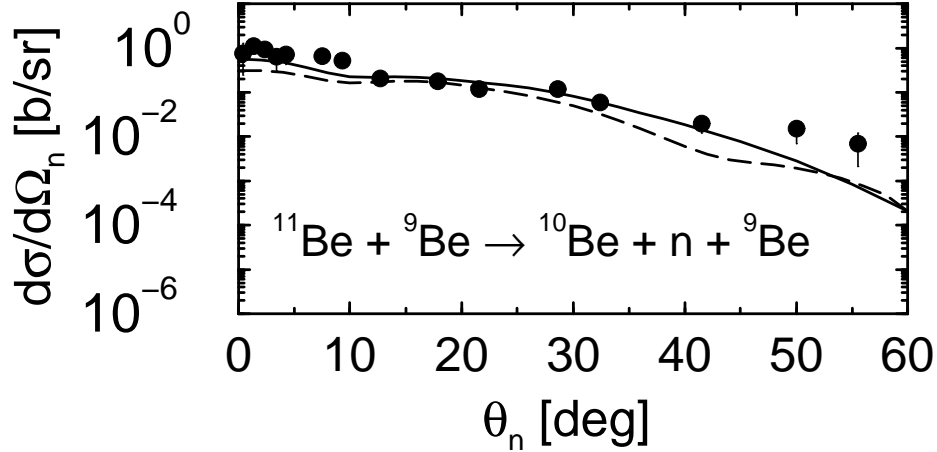,width=0.75\textwidth}}
\end{center}
\caption {The neutron angular distribution in the breakup of $^{11}$Be
on Be at 41 MeV/nucleon. The dashed line shows the calculation 
where a lower cut-off of 10 fm is applied in the 
$r_i$ integral [Eq. (\ref{eq18_5})] while
the solid line shows the result without any cut-off. The data are
taken from \cite{ann94}. 
}
\label{fig:figrm_5}
\end{figure}
Thus, almost $70{\%}$ of this cross section is accounted for by
regions larger than 10 fm.
This is a consequence of the large spatial extent of   
$^{11}$Be, which allows it to interact with the target nucleus 
even at a larger distance. 

\subsection{Parallel momentum distribution}
\begin{figure}[ht]
\begin{center}
\mbox{\epsfig{file=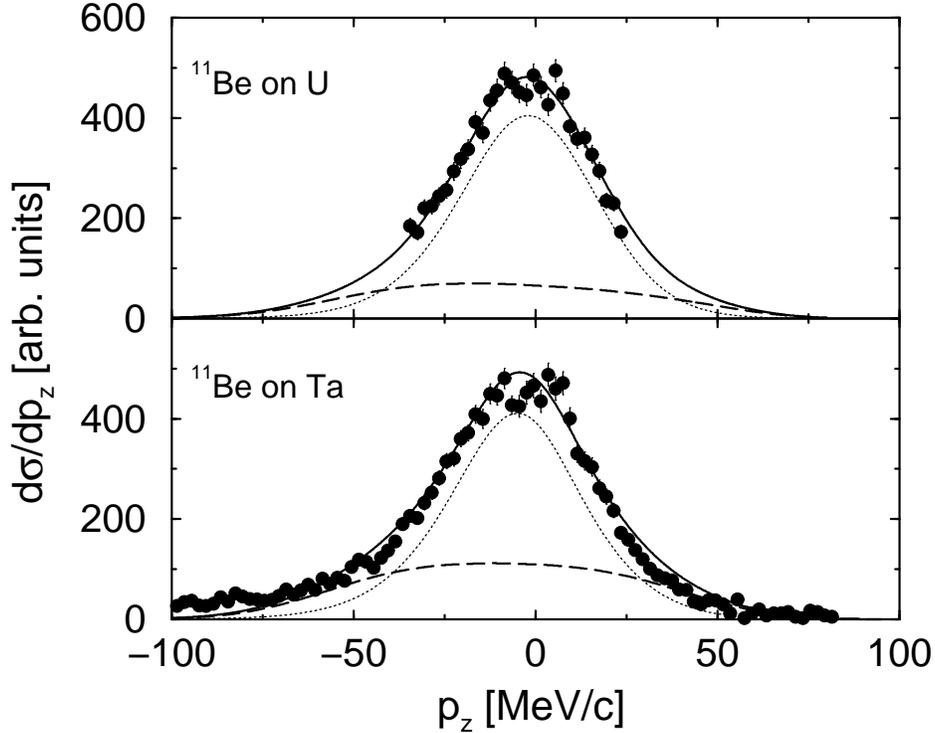,width=0.75\textwidth}}
\end{center}
\caption {
The parallel momentum distribution of the core
 in the breakup of $^{11}$Be
on U and Ta targets, at 63 MeV/nucleon beam energy, in the rest
frame of the projectile. The dotted and dashed
lines represent the pure Coulomb and nuclear breakup 
contributions, respectively while their coherent sums are 
shown by solid lines. The data are taken from~\protect\cite{kelly95_5}. 
}
\label{fig:figd_5}
\end{figure}

The parallel momentum distributions (PMDs) of the $^{10}$Be fragment
in the breakup of $^{11}$Be
on U and Ta targets, at 63 MeV/nucleon beam energy are presented in the rest
frame of the projectile, in Fig. 4. 
The core transverse momentum has been integrated from 0-500 MeV/c
and the neutron angle ($\theta_n$) has been integrated from $0^{\circ}$ to 
$30^{\circ}$. 
We have used the following sets of optical potentials here: 
$V_r=200~{\rm MeV},~r_r=1.23~{\rm fm},~a_r=0.9~{\rm fm},
~W_i=76.2~{\rm MeV},~r_i=1.49~{\rm fm},~a_i=0.38~{\rm fm}$ for the U
target and $V_r=200~{\rm MeV},~r_r=1.26~{\rm fm},~a_r=0.9~{\rm fm},
~W_i=76.2~{\rm MeV},~r_i=1.53~{\rm fm},~a_i=0.38~{\rm fm}$ for the Ta target.
The dotted and dashed lines show the contributions of the pure Coulomb
and nuclear breakups, respectively, while their coherent sums are represented by
solid lines. The coherent sum is normalized to the peak of the data,
which are given in arbitrary units, and the same normalization factor 
has been used for the pure Coulomb and pure nuclear contributions.

It is seen that around the peak region, the Coulomb contributions dominate.
This is because most of the contributions in this region come from forward
angles, where Coulomb breakup is the predominant mode. 
However, in the wings of the distribution (beyond about $|p_z| = 40$ MeV/c),
contributions come from large scattering angles and consequently the 
pure nuclear breakup dominates, in this region.

The FWHM of the distributions on U and Ta targets are found to be
48 MeV/c and 49 MeV/c, respectively. The width of the fragment momentum
distribution can be qualitatively related to the radial extent
of the coordinate space nuclear wavefunction of the projectile via
Heisenberg's uncertainty principle. Thus a narrow PMD width implies a large
spatial extension of the nuclear wavefunction in the coordinate space
for $^{11}$Be. 

\begin{figure}[ht]
\begin{center}
\mbox{\epsfig{file=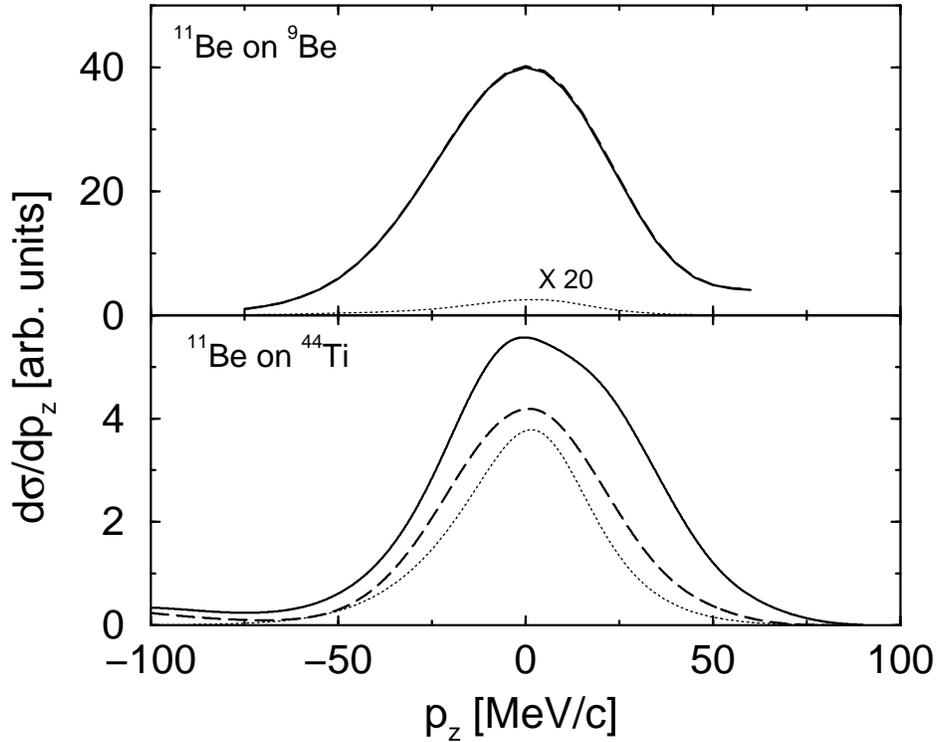,width=0.75\textwidth}}
\end{center}
\caption {
The parallel momentum distribution of the core
 in the breakup of $^{11}$Be
on $^{9}$Be and $^{44}$Ti targets, at 63 MeV/nucleon beam energy, in the rest
frame of the projectile. The dotted and dashed
lines represent the pure Coulomb and nuclear breakup 
contributions, respectively while their coherent sums are 
shown by solid lines. The pure Coulomb contribution in case of breakup
on a $^{9}$Be target is multiplied by a factor of 20 to make it visible. 
}
\label{fig:figd_5x}
\end{figure}
Our calculations with medium mass $^{44}$Ti and light
$^{9}$Be targets also leads us to similar conclusions. In Fig. 5
we present the PMDs of the $^{10}$Be fragment in the breakup of $^{11}$Be
on $^{44}$Ti and $^{9}$Be targets, at 63 MeV/nucleon beam energy, in the
rest frame of the projectile. The integrations over the transverse 
momentum and the neutron angle have been performed over the same range 
as in the case of heavy targets. The dotted and the dashed lines are the
pure Coulomb and pure nuclear contributions, respectively, while their
coherent sums are represented by solid lines. In case of breakup 
on a medium mass target ($^{44}$Ti) [Fig. 5 (lower panel)] the
pure Coulomb and pure nuclear contributions have nearly equal contributions,
as has been seen while calculating the relative energy spectra of the fragments.
Breakup on a light target ($^{9}$Be) [Fig. 5 (upper panel)] is expectedly 
pure nuclear dominated. The pure Coulomb contribution is extremely small
and is in fact multiplied by 20 to make it visible. The pure nuclear (dashed
line) and the coherent sum (solid line) of the pure nuclear and pure Coulomb
contributions thus almost coincide with each other. 
The FWHM of the distributions on $^{44}$Ti and $^{9}$Be targets are 
found to be 60 MeV/c and 56 MeV/c, respectively. They are in line with
the low widths mentioned earlier with heavy targets. 

\subsection{Total one-neutron removal cross section}
\begin{table}[ht]
\caption{Total one-neutron removal cross section, various contributions
from pure Coulomb and pure nuclear breakups, and their incoherent
sum for $^{11}$Be breakup 
on Au, Ti and Be targets, at beam energy of 41 MeV/nucleon.}
\begin{center}
\begin{tabular}{|c|c|c|c|c|c|c|}
\hline
Target &  Total & Pure Coulomb & Pure nuclear & Incoherent sum& 
    Expt.~\cite{ann94}\\
       & (b)    & (b)          & (b)          & (b)           &
    (b)               \\
\hline
$^{197}$Au & 2.10 & 1.88 & 0.21 & 2.09 & $2.5 \pm 0.5$  \\
$^{44}$Ti  & 0.403 & 0.177 & 0.189 & 0.37 & $0.55 \pm 0.11$  \\
$^{9}$Be   & 0.189 & 0.006 & 0.181 & 0.187 & $0.24 \pm 0.05$  \\
\hline
\end{tabular}
\end{center}
\end{table}

In Table II, we show the
contributions of pure Coulomb and pure nuclear breakup mechanisms to
the total one-neutron removal cross sections in the breakup of
$^{11}$Be on Au, Ti and Be targets at the beam energy of 41 MeV/nucleon.
The incoherent sum is obtained by simply adding the pure
Coulomb and pure nuclear cross sections.

For the heavy mass, high-Z target (Au) case it is seen that pure
Coulomb breakup accounts for about 90{\%} of the total cross section.
On a medium mass, medium-Z target (Ti), the pure Coulomb and nuclear
contributions to the total cross section, are nearly equal to each other,
while for the low mass, low-Z target (Be), the pure nuclear contribution
accounts for almost all of the total cross section.

The total one-neutron removal cross section on Au and Be targets 
does not seem to be affected by the CNI terms, while for the Ti
target case the incoherent sum seems to be about 10{\%} less than
the total cross section. Thus, it seems that the CNI terms 
manifests themselves more explicitly in more exclusive measurements,
like double differential cross sections than in
quantities like total cross sections.

\section{Summary and Conclusions}

In this paper we have presented an extended version of a fully 
quantum mechanical theory of halo breakup reactions \cite{cha02}
within the framework of post-form DWBA,
where the pure Coulomb, pure nuclear as well as their
interference terms are treated consistently within the same framework.
In this theory, both the Coulomb and nuclear interactions between 
the projectile and the target nucleus are treated to all orders, but 
the fragment-fragment interaction is treated in the first order.
The full ground state wavefunction of the projectile corresponding to
any orbital angular momentum structure enters as an input to this theory.
The lack of proper knowledge of appropriate optical potentials, particularly
in the halo projectile-target channel is a source of uncertainty in 
the calculations. However this is the case for all reaction studies of
halo nuclei where distorted waves in the projectile-target channel are used,
as has already been pointed out in Ref. \cite{cha02}.

We applied our theory to study the breakup of one-neutron halo nucleus
$^{11}$Be on several targets. 
Results for the neutron energy distribution in the breakup
of $^{11}$Be on Au at 41 MeV/nucleon emphasize the fact that the
Coulomb-nuclear interference terms are both energy and angle dependent. 
They are almost of the same magnitude as the nuclear breakup 
contributions and this leads to a difference in the coherent and incoherent
sums of the Coulomb and nuclear terms, more so at forward angles. 
The parallel momentum distribution of the $^{10}$Be fragment in the breakup
reaction of $^{11}$Be in the Coulomb and nuclear fields of U and Ta targets
have also been calculated at 63 MeV/nucleon. It is seen that the
region around the peak of the distribution, which gets 
substantial contributions from forward scattered fragments, is Coulomb 
dominated, while in the wings of the distribution, where contributions
come from fragments scattered at large angles, the nuclear breakup 
contributions dominate. The FWHMs of the distributions were also found to
be small consistent with the expectation of the wavefunction of $^{11}$Be 
to have a large spatial extent in the coordinate space. Parallel momentum
distributions in the breakup on medium mass (Ti) and light (Be) targets 
also confirmed a small FWHM. The relative energy spectra of the fragments
(neutron and $^{10}$Be) emitted in the breakup of $^{11}$Be on 
Pb, Ti and C targets, at the beam energy of 72 MeV/nucleon have also
been calculated.
While the breakup on the light target was highly nuclear dominated, that
on a heavy target required the nuclear and the CNI
terms for a better explanation of the data particularly at 
higher relative energies. In case of breakup on a medium mass target, 
the total pure Coulomb and pure nuclear contributions were nearly equal
in magnitude. The CNI terms were found to have little 
impact on the total one-neutron removal cross section in the breakup of
$^{11}$Be on heavy and light targets at the beam energy of 41 MeV/nucleon,
but on a medium mass (Ti) target the CNI terms
were almost 10${\%}$ of the total one-neutron removal cross section.
Thus in many sophisticated experiments planned in the future one has to 
look into the role played by the CNI terms in 
analyzing the experimental data.

The full quantal theory of one-neutron halo
breakup reactions, presented in this paper, can  be applied to
describe the $(a,b\gamma)$ reaction provided the inelastic breakup
mode is also calculated within this theory. These studies are in 
progress. 
There is also a need to extend the theory to describe the halo
breakup at higher beam energies for which data have been taken
at GSI, Darmstadt. This can be achieved by introducing the
eikonal expansion of the distorted waves, instead of the partial
wave expansion as done here.

\acknowledgments
It is a pleasure to thank Prof. R. Shyam for many illuminating discussions
and for going through the manuscript. Thanks, also to Prof. T. Nakamura
for providing the experimental data shown in the bottom panel of Fig. 2
in a tabular format.
\appendix

\section{Validity of the local momentum approximation}

The LMA provides a way of taking into
account the finite range effects in the DWBA theory. It leads to 
the factorization of the breakup amplitude [Eq. (\ref{eq2_5})],
which makes its numerical calculation relatively simpler.
As discussed in Ref. \cite{shy85}, a condition of validity
of the LMA is that the quantity
\begin{eqnarray}
\eta(r) = {1\over 2}K(r)|dK(r)/dr|^{-1}
\end{eqnarray}
evaluated at a representative distance $R$ should be larger than the
projectile radius ($r_a$).
\begin{figure}[ht]
\begin{center}
\mbox{\epsfig{file=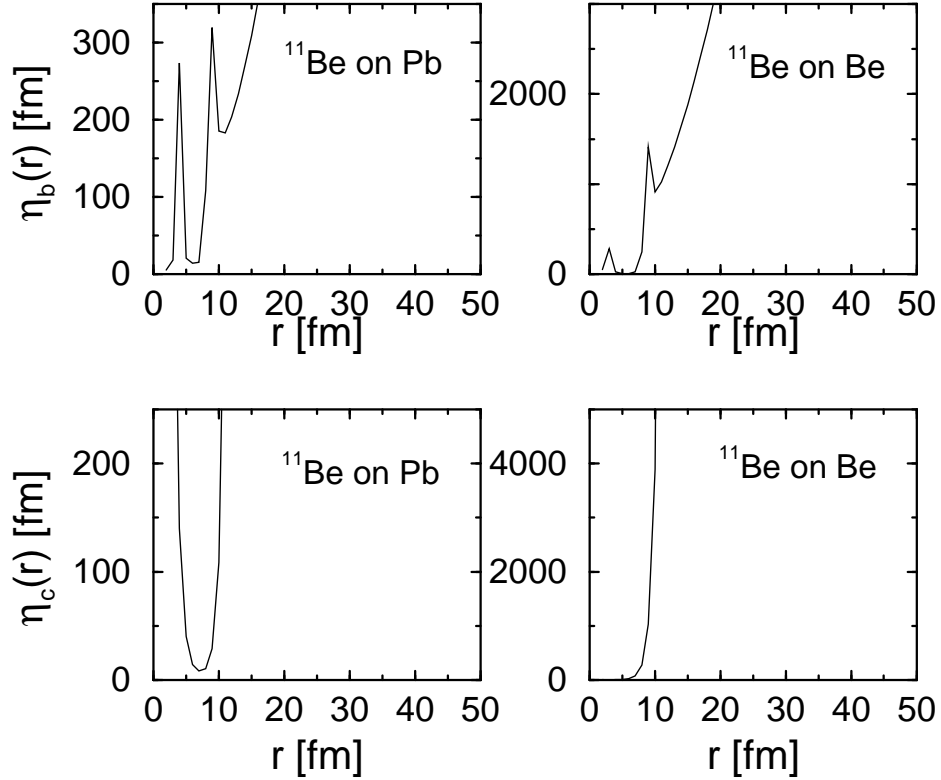,width=.75\textwidth}}
\end{center}
\caption{Variation of $\eta_b(r)$ (top half) and $\eta_c(r)$
(bottom half) with $r$, for the breakup of $^{11}$Be on Pb, 
at 72 MeV/nucleon (left half) and
$^{11}$Be on Be at 41 MeV/nucleon (right half).
}
\label{fig:figou}
\end{figure}
In our case the LMA is done on both the outgoing fragments ($b$ and $c$). 
In Fig. 6, we show the variation of $\eta_b(r)$ (top half) and $\eta_c(r)$
(bottom half) with $r$, 
for the breakup of $^{11}$Be on a Pb target, at the beam energy of 
72 MeV/nucleon (left half) and
$^{11}$Be on a Be target at the beam energy of 
41 MeV/nucleon (right half), where both Coulomb and nuclear potentials
are included in the term $V_j(r)$ in the definition of $K(r)$ [Eq. (\ref{ea6})].
At $r = 10$ fm,
$\eta_b(r) = 185$ fm and $\eta_c(r)= 108$ fm for $^{11}$Be incident on a
Pb target and $\eta_b(r) = 914$ fm and $\eta_c(r)= 3892$ fm for 
$^{11}$Be incident on a Be target. These values are much larger than the
rms radius of $^{11}$Be, which is about 2.91 fm. (Incidentally, the spikes 
and turns in $\eta(r)$ at low $r$ are due to the presence of the short
ranged nuclear potential.)

\begin{figure}[ht]
\begin{center}
\mbox{\epsfig{file=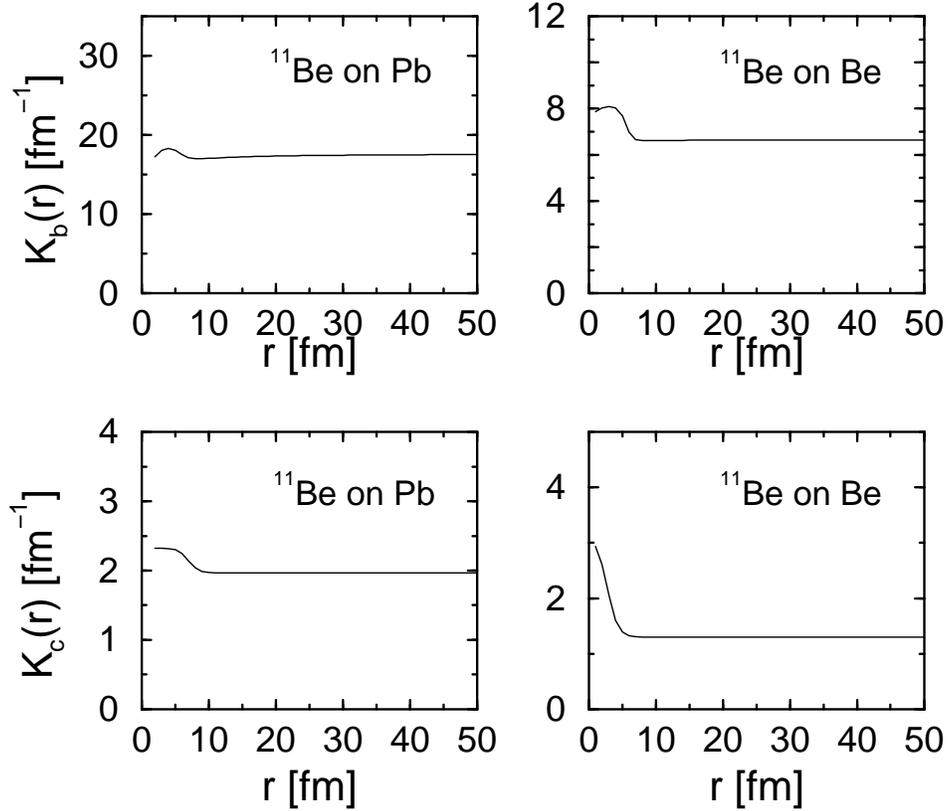,width=.75\textwidth}}
\end{center}
\caption{Variation of $K_b(r)$ (top half) and $K_c(r)$
(bottom half) with $r$, for the breakup of $^{11}$Be on Pb, 
at 72 MeV/nucleon (left half) and
$^{11}$Be on Be at 41 MeV/nucleon (right half).
}
\label{fig:figov}
\end{figure}

The variations of $K_b(r)$ and $K_c(r)$, 
the magnitudes of the local momentum (LM), with $r$ for the above 
mentioned reactions are 
shown in Fig. 7. We see that $K_b(r)$ and $K_c(r)$ 
remains practically constant
for $r > 8$ fm. Due to the peripheral nature of breakup reactions,
this region contributes maximum to the cross section. 
Therefore, our choice of a constant magnitude
for the local momentum evaluated at 10 fm is well justified. 
Thus the condition of validity of the LMA is 
well fulfilled in these cases.

We have also performed calculations for different 
LM directions of $b$ and $c$. We denote the different
combination of directions as $D_1$: both the LM angles
of $b$ and $c$
are taken along asymptotic directions, $D_2$: the LM angles of $b$ are
taken to be zero while the LM angles of $c$ are taken along 
the asymptotic direction, $D_3$: the LM angles of $b$ are
taken to be half those of the asymptotic direction while the LM 
angles of $c$ are taken along the asymptotic direction,
$D_4$: the LM angles of $b$ are
taken along the asymptotic direction while the LM angles of $c$ are taken
to be zero, 
$D_5$: the LM angles of $b$ are
taken along the asymptotic direction while the LM angles of $c$ are taken
to be half those of the asymptotic direction, 
$D_6$: both the LM angles of $b$ and $c$
are taken to be zero, and
$D_7$: both the LM angles of $b$ and $c$
are taken to be half those of asymptotic directions. 
\begin{table}[ht]
\begin{center}
\caption[T4]{Calculated value of the total one-neutron
removal cross section for $^{11}$Be on Au and Be targets at 41 MeV/nucleon
for different local momentum directions (see text). 
}
\vspace{0.5cm}
\begin{tabular}{|cccccccc|}
\hline
Projectile & $D_1$ & $D_2$& $D_3$ &  $D_4$& $D_5$&
$D_6$ &$D_7$\\
+ target& (b)& (b)&(b) & (b)& (b) & 
(b) &(b)\
\\ \hline
$^{11}$Be~+~Au & 2.10 & 2.13 & 2.14 & 2.16 & 2.10 & 2.24 & 2.29 \\ 
$^{11}$Be~+~Be & 0.189 & 0.204 & 0.192 & 0.167 & 0.162 & 0.141 & 0.150 \\ \hline
\end{tabular}
\end{center}
\end{table}
In Table III, we present the total one-neutron removal cross section
in the breakup of $^{11}$Be on Au and Be targets at 41 MeV/nucleon for
different directions of the LM.
We note that these cross sections depend
on the local momentum directions only to the extent of $10-15{\%}$.

\begin{figure}[ht]
\begin{center}
\mbox{\epsfig{file=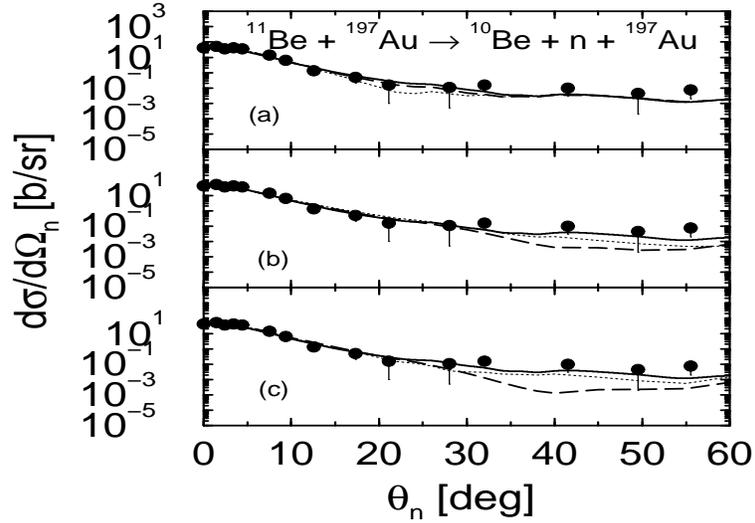,height=7.0cm,width=.60\textwidth}}
\end{center}
\caption{The neutron angular distribution for the breakup reaction
$^{11}$Be + Au $\to$ $^{10}$Be + $n$ + Au, at 41 MeV/nucleon for
different LM directions (see text).
}
\label{fig:figow}
\end{figure}
\begin{figure}[h]
\begin{center}
\mbox{\epsfig{file=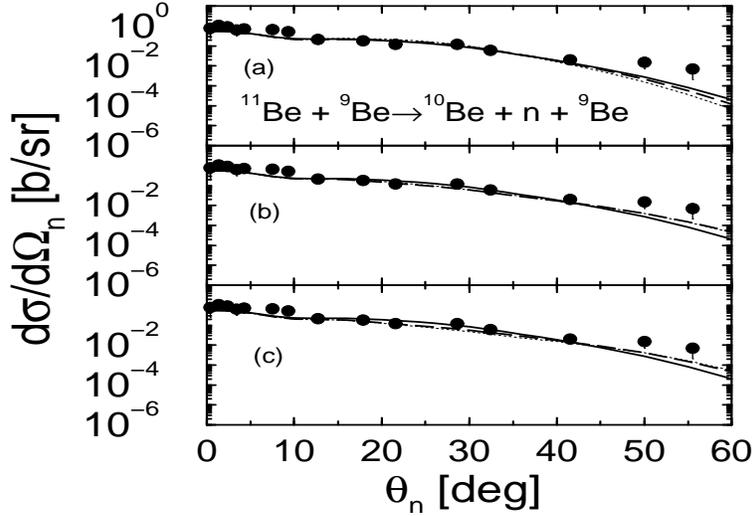,height=7.0cm,width=.60\textwidth}}
\end{center}
\caption{The neutron angular distribution for the breakup reaction
$^{11}$Be + Be $\to$ $^{10}$Be + $n$ + Be, at 41 MeV/nucleon for
different LM directions (see text).
}
\label{fig:figox}
\end{figure}   

In Figs. 8 and 9, we show the variation of the neutron angular distribution
for different combinations of LM directions of the core and the 
valence neutron in
the breakup of $^{11}$Be on Au and Be targets, respectively, at 41 MeV/nucleon.
In both the figures, the solid line shows the calculation with direction $D_1$.
In panel (a), of both Figs. 8 and 9, calculations are shown for
directions $D_2$ (dotted line) and $D_3$ (dashed line), panel (b) shows
it for $D_4$ (dotted line) and $D_5$ (dashed line), while panel (c)
shows the same for $D_6$ (dotted line) and $D_7$ (dashed line).
We see that for the Be target case the neutron angular distributions are 
minimally dependent on various LM directions, but for the Au target case,
some dependence is observed at large angles. 
However, the magnitudes of
the cross sections here are very small. Consequently no such angular
dependence will be observed
in cross sections which involve angular integrations for both 
the outgoing fragments or in more 
inclusive cross sections. 

 
\end{document}